\documentclass[conference]{IEEEtran}
\IEEEoverridecommandlockouts
\usepackage{cite}
\usepackage{array}
\usepackage{amsmath,amssymb,amsfonts}
\usepackage{algorithmic}
\usepackage{graphicx}
\usepackage{textcomp}
\usepackage{xcolor}
\def\BibTeX{{\rm B\kern-.05em{\sc i\kern-.025em b}\kern-.08em
    T\kern-.1667em\lower.7ex\hbox{E}\kern-.125emX}}
\begin{document}

\title{Neural Network-Augmented Iterative Learning
Control for Friction Compensation of Motion Control Systems with Varying
Disturbances\\
}

\author{\IEEEauthorblockN{1\textsuperscript{st} Ali Mashhadireza}
\IEEEauthorblockA{\textit{Department of Mechanical and Industrial Engineering} \\
\textit{Northeastern University}\\
Boston, United States \\
mashhadireza.a@northeastern.edu}
\and
\IEEEauthorblockN{2\textsuperscript{nd} Ali Sadighi}
\IEEEauthorblockA{\textit{Department of Mechanical Engineering} \\
\textit{University of Tehran}\\
Tehran, Iran \\
asadighi@ut.ac.ir}
\and
}

\maketitle

\abstract{This paper proposes a robust control strategy that integrates Iterative Learning Control (ILC) with a simple lateral neural network to enhance the trajectory tracking performance of a linear Lorentz force actuator under friction and model uncertainties. The ILC compensates for nonlinear friction effects, while the neural network estimates the nonlinear ILC effort for varying reference commands. By dynamically adjusting the ILC effort, the method adapts to time-varying friction, reduces errors at reference changes, and accelerates convergence. Compared to previous approaches using complex neural networks, this method simplifies online training and implementation, making it practical for real-time applications. Experimental results confirm its effectiveness in achieving precise tracking across multiple tasks with different reference trajectories.
}



\maketitle

\section{Introduction}\label{sec1}

Precise trajectory tracking and adherence to desired motion profiles are essential in a wide range of industrial applications. However, achieving such precision is often challenged by practical issues, including system imperfections, model uncertainties, measurement noise, and nonlinearities such as friction. Among these, friction presents a particularly complex and variable phenomenon, making its accurate compensation difficult.\\

Various methods have been proposed to identify nonlinearities and the dynamics of plants in different applications~\cite{Rahmat2012, Richer2000, Mashhadireza2022, Mashhadireza2025-nd}. Effectively compensating for friction typically demands robust control strategies. Traditional techniques often involve applying forces to counteract friction, but they require precise knowledge of frictional forces, which may change with time, position, or velocity. Common model-based approaches include relay-based identification of Coulomb, viscous, and Stribeck friction components, followed by their integration into feedforward controllers~\cite{Chen2009, Huang2019, Friedland1992}. While effective in some cases, these methods may struggle when friction is highly variable or uncertain.\\

Model-free strategies have emerged as promising alternatives. Soft computing techniques such as adaptive control~\cite{Misovec1998}, fuzzy logic~\cite{Mostefai2009}, and neural networks~\cite{Tan2002, Scholl2024, Ciliz2007} offer viable solutions. Adaptive controllers, for instance, adjust parameters dynamically to accommodate changing friction characteristics, although they often require prior knowledge of friction behavior and may be complex to implement~\cite{Huang2019, Bona2010}. Such limitations can constrain their practical applicability, especially in dynamic environments.\\

Iterative Learning Control (ILC) offers an effective approach for mitigating friction effects by refining control efforts across repeated tasks based on past iterations~\cite{Spong2006}. ILC has demonstrated strong potential for improving both tracking performance and robustness and has been explored as a tool for friction compensation~\cite{Lee2019, Norrlöf2020, Riaz2023}. Integrating ILC with neural networks enables the capture of complex friction patterns that traditional methods may fail to model accurately.\\

The combination of ILC and neural networks addresses challenges associated with nonlinearities and variable friction, particularly for tasks with completely varying setpoints. Xu et al.~\cite{Xu2022} proposed a learning-based ILC where neural networks adaptively tune ILC gains, showing promise but heavily depending on parameter tuning. Zhang et al.~\cite{Zhang2021} introduced a switching neural network ILC to handle highly nonlinear friction, but this approach excluded ILC from the learning loop, potentially reducing effectiveness and increasing computational costs during online training.\\

Additionally, research in~\cite{Patan2018} has combined neural networks with ILC to improve convergence and transient response, though it did not address scenarios involving changing reference commands. Prior studies have applied ILC to handle small variations in reference signals~\cite{Boeren2016}, but leveraging neural networks to extend ILC across different task scenarios and varying setpoints remains underexplored. This underscores the need for more adaptive and efficient strategies for dynamic environments.\\

In this work, we apply ILC together with a simply structured neural network to an electromagnetic linear Lorentz actuator for enhanced motion control~\cite{Mashhadireza2023b, Mashhadireza2024-mk, Saeedi2021, Sarvjahny2023, Erfanimatin2024}. The approach tackles model uncertainties, including position- and current-dependent force constants, and friction effects using a model-based ILC. A Kalman filter is incorporated to improve robustness, and the feedforward signal from ILC is combined with the neural network output to capture friction dynamics. This hybrid scheme enhances ILC robustness, reduces convergence time, and adapts effectively to varying tasks.\\

By integrating ILC with neural network-based compensation, the proposed method provides a robust and efficient solution for motion control in the presence of complex friction and changing setpoints. Assuming time-varying friction effects, the approach adapts by updating the ILC effort in real-time alongside the neural network. Unlike previous works that rely on complex networks and remove ILC from learning, our method uses a simpler network structure, facilitating practical online training in conjunction with ILC.\\

The paper is organized as follows: Section II details the integration of ILC with neural networks. Section III describes the experimental setup and system modeling. Sections IV and V present simulation and experimental results, demonstrating the effectiveness of the proposed method under varying conditions. Section VI concludes with key findings and future research directions.

\section{Integration of Iterative Learning Control and Neural Networks}\label{sec2}

This section presents the theoretical background of Iterative Learning Control (ILC) and its integration with neural networks. ILC is an effective approach for improving the performance of control systems that execute repetitive tasks, by learning from previous iterations to progressively reduce tracking errors.

\begin{figure}[htbp]
\centerline{\includegraphics[width=\columnwidth]{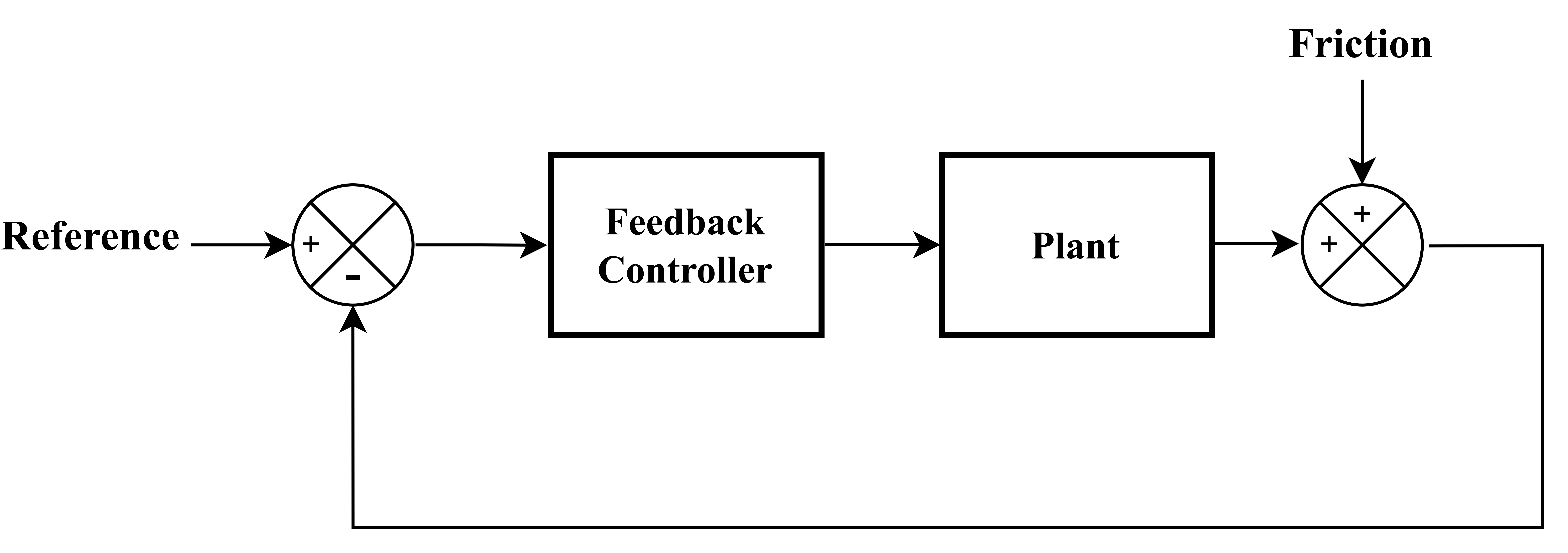}}
\caption{Sample figure illustrating the ILC framework.\label{Fig.1}}
\end{figure}

\noindent
Based on the relations depicted in \textbf{Fig.~\ref{Fig.1}}, the plant output can be expressed as:
\begin{equation}
y_j(z) = G(z) \left( d_j'(z) + u_j(z) \right) = G(z) u_j(z) + d_j(z)
\label{Eq.1}
\end{equation}

\noindent
Here, $j$ denotes the iteration index, $G(z)$ represents the causal plant dynamics, and $y_j$, $u_j$, and $d_j$ correspond to the output, control input, and disturbances at iteration $j$, respectively. The disturbances $d_j$ are decomposed into repeated and non-repeated components:
\begin{equation}
d_j = d_{(r,j)} + d_{(n,j)}
\label{Eq.2}
\end{equation}
where $d_{(r,j)}$ are persistent across iterations, and $d_{(n,j)}$ represent random variations such as noise and external perturbations.

\subsection{Iterative Learning Control Algorithm}\label{subsec21}
\begin{figure}[htbp]
\centerline{\includegraphics[width=\columnwidth]{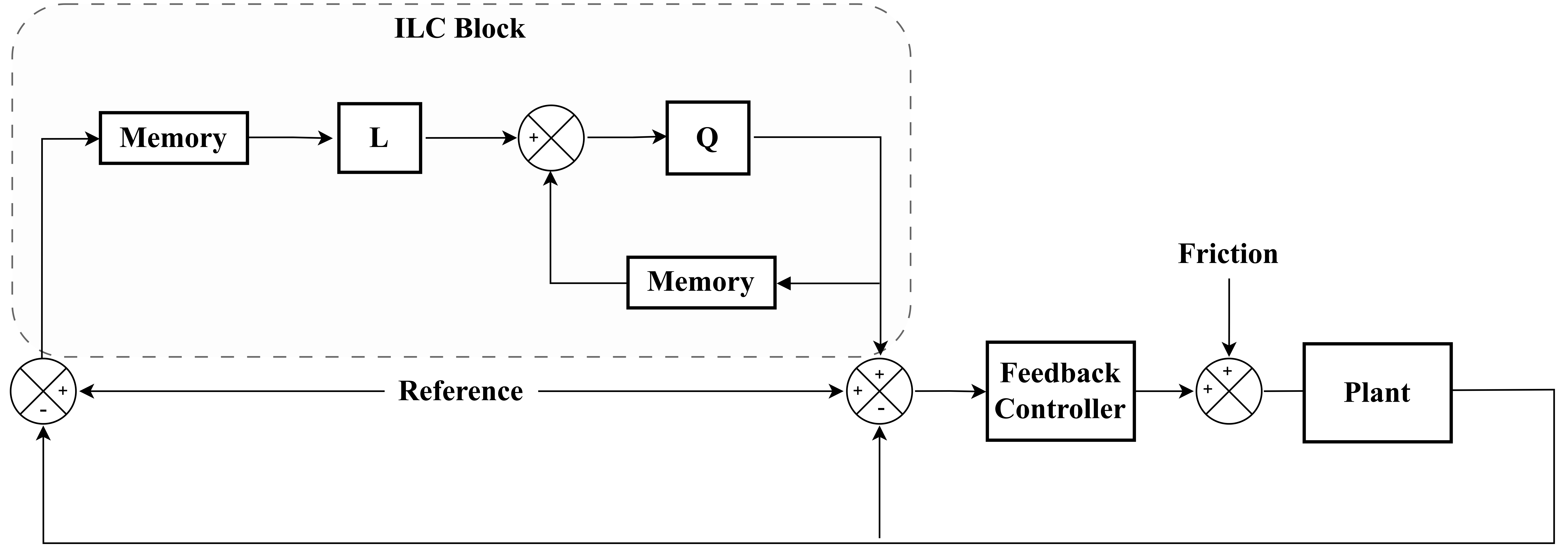}}
\caption{Learning dynamics of ILC.\label{Fig.2}}
\end{figure}

A standard ILC update law can be written as~\cite{Spong2006}:
\begin{equation}
u_{(j+1)}(z) = Q(z) \left(u_j(z) + L(z) e_j(z)\right)
\label{Eq.3}
\end{equation}
where $Q(z)$ is a filter that shapes the learning process and attenuates high-frequency components for stability, and $L(z)$ is the learning function that determines how the control error 
\begin{equation}
e_j(z) = r(z) - y_j(z)
\label{Eq.4}
\end{equation}
is incorporated at iteration $j$. Figure~\ref{Fig.2} illustrates a basic ILC configuration.

\noindent
The disturbance $d_j[k]$ significantly influences ILC performance. By mitigating the non-repeated component $d_{n,j}$ using conventional feedback (e.g., PID) and a low-pass $Q$-filter, the algorithm can primarily target repeated disturbances $d_{r,j}$, which are the primary focus of ILC compensation.

The learning function $L(z)$ can be selected as the inverse of the estimated plant $\hat{G}$, providing fast convergence but relying on model accuracy. In this case, (\ref{Eq.3}) becomes:
\begin{equation}
u_{j+1}(z) = Q(z)\left(u_j(z) + \beta \hat{G}^{-1} e_j(z) \right)
\label{Eq.5}
\end{equation}
where $\beta$ is a learning rate that ensures robustness and convergence. Monotonic trial-to-trial convergence is guaranteed if~\cite{Harte2005}:
\begin{equation}
\sup_{|z|=1} \left|\frac{1}{\beta} - \hat{G}^{-1} G(z)\right| < \left|\frac{1}{\beta}\right|
\label{Eq.6}
\end{equation}

\noindent
In this study, we focus on integrating a neural network alongside ILC to handle multiple reference commands. As the reference changes, friction dynamics vary, challenging the standard ILC assumption. By leveraging a neural network to predict the converged ILC effort corresponding to each reference, the number of iterations required for convergence after a reference change is reduced, while maintaining system stability.

\subsection{Incorporation of Neural Networks in ILC}\label{subsec22}
At steady state, the learned ILC effort $u_\infty$ satisfies:
\begin{equation}
u_\infty = u_\infty + \hat{G}^{-1} e_\infty = u_\infty + \hat{G}^{-1} (r - y_\infty)
\label{Eq.7}
\end{equation}
Given $y_\infty = G(y_d + u_\infty) + d$, it follows that:
\begin{equation}
u_\infty = (\hat{G}^{-1} - 1) y_d + d = u_l + u_n
\label{Eq.9}
\end{equation}
where $u_l$ is the linear portion representing feedforward compensation, and $u_n$ is the nonlinear component capturing model mismatches and complex friction effects~\cite{Norrlöf2020}.

\noindent
The nonlinear portion $u_n$ is iteratively learned and is unique for each reference command. By training a neural network to predict $u_n$ based on position and velocity of the reference trajectory, the system can preemptively compensate nonlinearities, reducing the number of iterations needed for ILC convergence in multi-task scenarios.

\subsection{Kalman Filter Implementation for Robust Performance}\label{subsec23}
\begin{figure}[htbp]
\centerline{\includegraphics[width=\columnwidth]{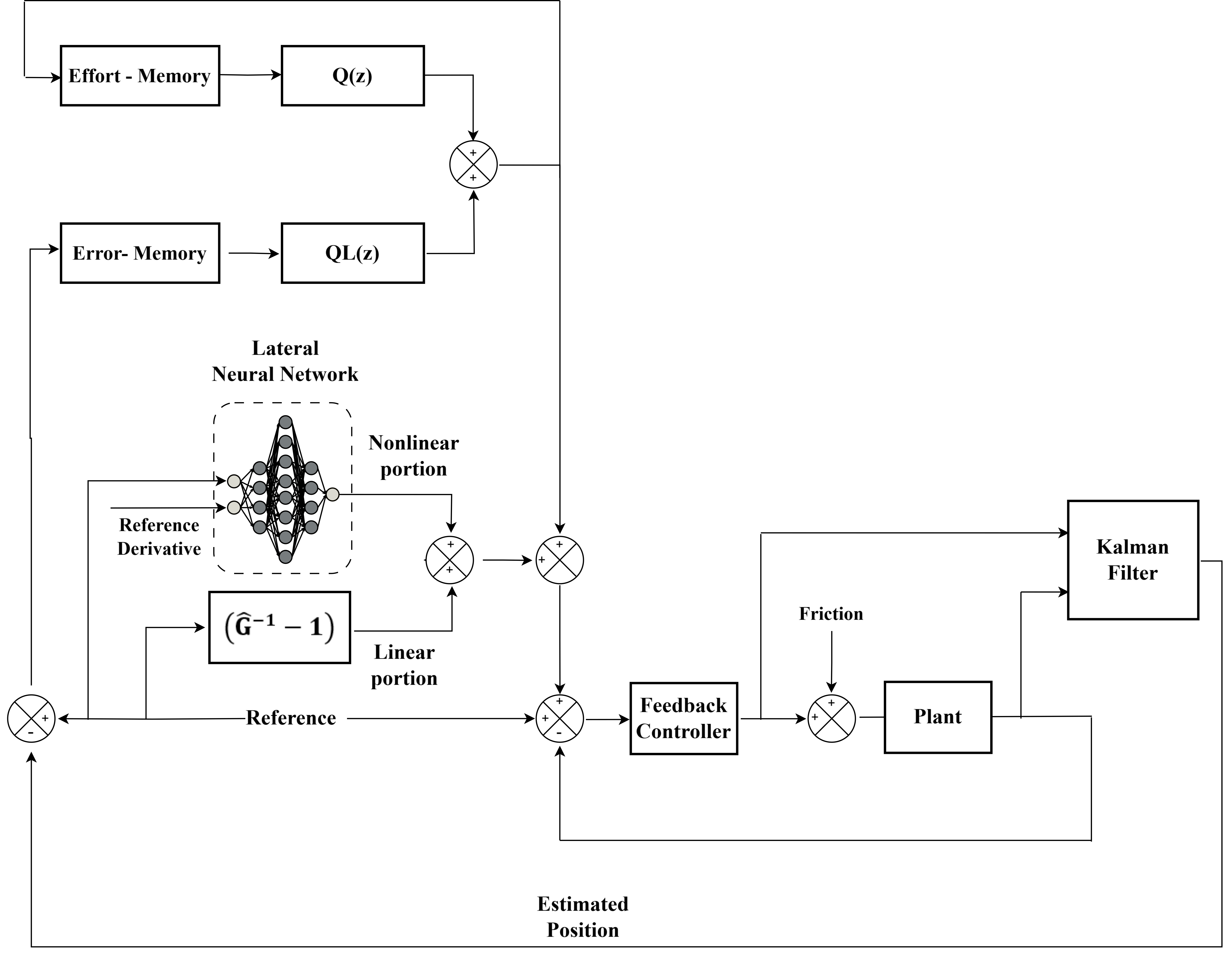}}
\caption{Integration of ILC and neural network with Kalman filtering.\label{Fig.3}}
\end{figure}

High-bandwidth $Q$-filters improve convergence but amplify noise. To address this, an online Kalman filter estimates position recursively, providing a cleaner signal for ILC, allowing robust performance with faster convergence. The integrated control schematic is shown in \textbf{Fig.~\ref{Fig.3}}.

\subsection{Stability Analysis}\label{subsec24}
The stability of neural network-based ILC relies on the boundedness of the neural network output. Error dynamics can be expressed as:
\begin{equation}
E_{j+1} = (R(z) - D)(1 - Q(z)) + Q(z)(1 - GL(z)) E_j(z)
\label{Eq.11}
\end{equation}
and the necessary and sufficient condition for trial-to-trial convergence is:
\begin{equation}
\frac{E_{\infty}(z) - E_{i+1}(z)}{E_{\infty}(z) - E_i(z)} = Q(z)(1 - LG(z))
\label{Eq.13}
\end{equation}
ensuring convergence as long as:
\begin{equation}
\sup_{|z|=1} |1 - LG(z)| < \sup_{|z|=1} \left|\frac{1}{Q(z)}\right|
\label{Eq.14}
\end{equation}
Thus, bounded neural network outputs guarantee that the monotonic trial-to-trial convergence of the system is preserved.

\section{Experimental Setup}\label{sec3}

Electrical actuators, such as voice coil actuators, are particularly sensitive to parameter uncertainties and nonlinearities, including variations in force constants, friction, and hysteresis, which can negatively affect performance.  

\begin{figure}[htbp]
\centerline{\includegraphics[width=\columnwidth]{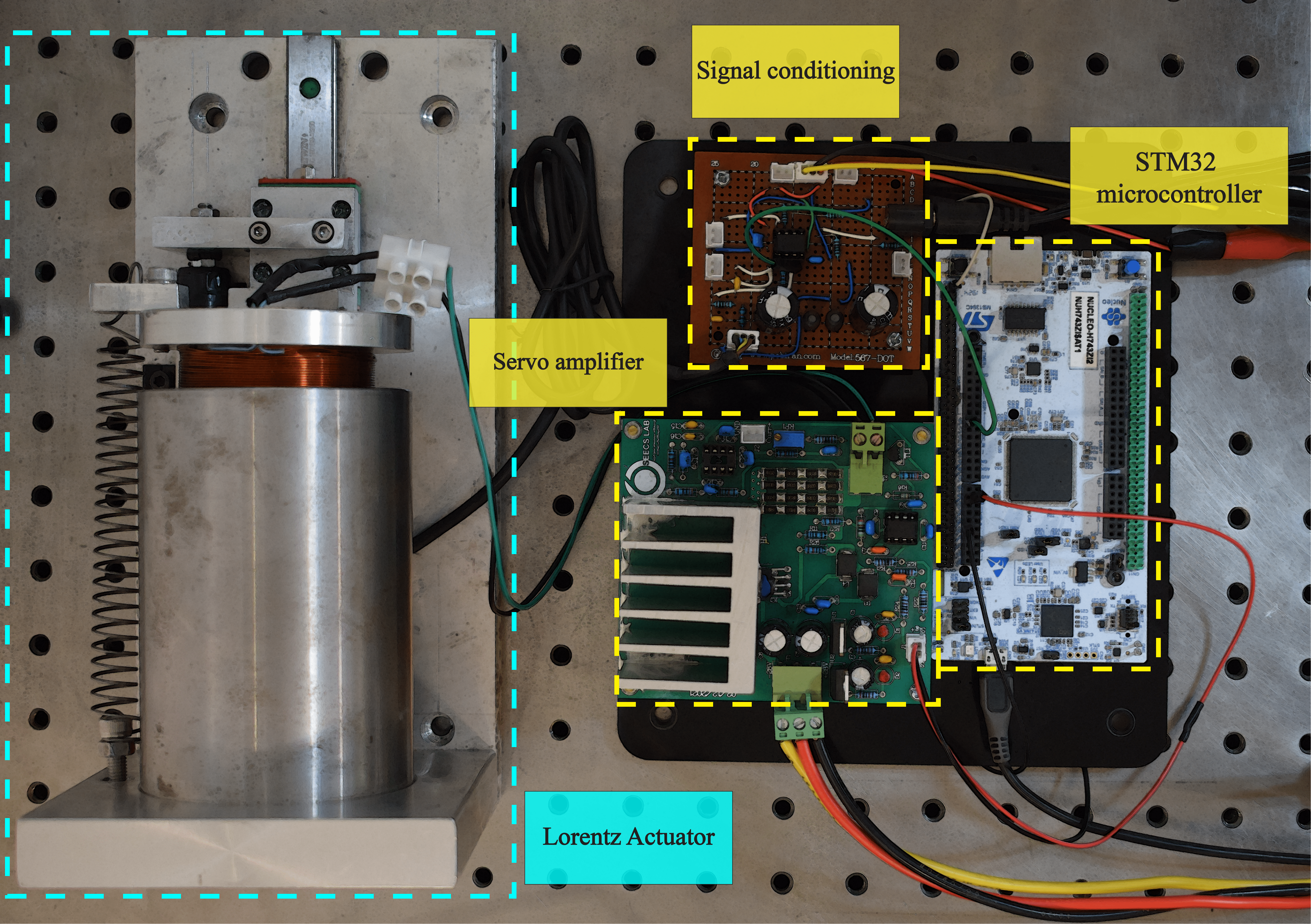}}
\caption{Mechanical and electrical components of the test setup.\label{Fig.4}}
\end{figure}

\noindent
The experimental setup employs an electromagnetic linear Lorentz actuator designed for precise trajectory tracking. The actuator consists of a permanent magnet and a coil, with the magnetic field oriented perpendicular to the coil current, generating an electromagnetic force according to Lorentz's law to drive motion.  

\noindent
As illustrated in \textbf{Fig.~\ref{Fig.4}}, the actuator is divided into mechanical and electrical sections. Mechanically, it includes a wound rotor and a stator holding the permanent magnet, with steel components providing structural stability and reduced friction. A spring enhances system stability, modeling the actuator as a mass-spring-damper system. Position feedback is obtained via a linear SLPT1005k1M potentiometer, which accurately measures rotor displacement.  

\noindent
Electrically, an STM32H743ZI2 microcontroller processes signals and controls the rotor position. Analog position signals are converted to digital form and transmitted as voltage commands to a servo amplifier, which drives the coil current using operational amplifiers for precise signal conditioning.  

\noindent
To improve measurement accuracy and reduce noise, the setup incorporates several filters. A CLC filter with a 100 Hz corner frequency suppresses high-frequency noise from the sensor supply, while a 500 Hz anti-aliasing filter prevents aliasing during data acquisition, ensuring signal integrity.  

\noindent
The system dynamics consider both mechanical and electrical components, with the mechanical dynamics dominating due to their lower bandwidth. The electrical portion is approximated as a gain converting voltage commands into current, which is then mapped into force through a constant force factor. The mechanical dynamics are modeled as a second-order system:
\begin{equation}
X(s) = \frac{K_t K_a}{Ms^2 + Cs + K_s} \quad V(s) = G(s)V(s)
\label{Eq.15}
\end{equation}
where $K_a$ is the servo amplifier gain, $K_t$ is the force

\section{Simulation}\label{sec4}

Prior to real-world implementation of the Iterative Learning Control approach, it is essential to verify the feasibility of the designed controllers through simulation. This ensures that feedback or feedforward controllers do not induce actuator saturation or integrator windup, which could introduce nonlinearities and compromise stability. Incorporating realistic constraints during simulation is crucial to confirm the practical applicability of the algorithm. Accordingly, the dynamical model of the experimental plant is used to simulate the complete set of control strategies.

\noindent
Simulations and practical implementations assume a duration of 6 seconds with a sampling interval of 0.001 seconds. The control signal is restricted to ±1.5 V to prevent degradation of the servo amplifier (OPA548T) and avoid excessive current that could damage the actuator's wound wires. Initially, a carefully tuned feedback controller is designed. The discrete-time Proportional-Integral (PI) controller is expressed as:

\begin{equation}
C(z) = 0.12 + 0.5 \times 10^{-3} \frac{1}{z-1}
\label{Eq.17}
\end{equation}

\noindent
Additionally, a white noise signal with variance matching that of the SLPT1005k1M potentiometer is included in the simulation.

\subsection{LuGre Friction Model}\label{subsec41}

To capture the frictional effects observed in the physical actuator, the LuGre model is employed. At the microscale, contact surfaces have asperities that behave as elastic bristles. Under tangential forces, these bristles deform and generate friction. When the applied force exceeds a threshold, some bristles slip, producing dynamic friction described by:

\begin{equation}
\frac{dy}{dt} = v - \frac{|v|}{g(v)} y
\label{Eq.18}
\end{equation}

\noindent
Here, \( g(v) \) models the Stribeck effect, representing the transition from stiction (\( F_s \)) to Coulomb friction (\( F_c \)) as velocity increases. The resulting friction force \( F \) is:

\begin{equation}
F = \sigma_0 y + \sigma_1 \frac{dy}{dt} + \sigma_2 v
\label{Eq.19}
\end{equation}

\noindent
where \( \sigma_0 \) is the bristle stiffness, \( \sigma_1 \) is the damping coefficient, and \( \sigma_2 \) represents viscous friction. The simulation parameters for the LuGre model are summarized in \textbf{Table~\ref{Tab1}}.

\begin{table}[ht]
\centering
\caption{LuGre model parameters used in simulation.}
\label{Tab1}
\renewcommand{\arraystretch}{1.3} 
\begin{tabular}{|c|c|}
\hline
\textbf{Parameter} & \textbf{Value} \\
\hline
$\sigma_0$ & 1067 \\
\hline
$\sigma_1$ & 1264911 \\
\hline
$\sigma_2$ & 0.7 \\
\hline
$F_c$ & 40 \\
\hline
$F_s$ & 60 \\
\hline
$v_s$ & 0.001 \\
\hline
\end{tabular}
\end{table}

\noindent
To improve steady-state performance, a Kalman filter is integrated with the ILC algorithm, enabling higher bandwidth in the Q-filter design. A fourth-order low-pass Butterworth filter is also applied, with transfer function:

\begin{equation}
Q(z) = 10^{-2} \frac{0.3 z^4 + z^3 + 2 z^2 + z + 0.3}{z^4 - 2.61 z^3 + 2.72 z^2 - 1.31 z + 0.24}
\label{Eq.20}
\end{equation}

\noindent
The Kalman filter design parameters include \(Q\) (process noise covariance), \(D\) (direct transmission matrix), \(R\) (measurement noise covariance), \(\vec{x}_0\) (initial state estimate), and \(P_0\) (initial error covariance). Values are provided in \textbf{Table~\ref{Tab2}}.

\begin{table}[ht]
\centering
\caption{Kalman filter design parameters.}
\label{Tab2}
\renewcommand{\arraystretch}{1.3} 
\begin{tabular}{|c|c|}
\hline
\textbf{Parameter} & \textbf{Value} \\
\hline
$Q$ & $\begin{pmatrix} 10^{-5} & 0 \\ 0 & 10^4 \end{pmatrix}$ \\
\hline
$\vec{x}_0$ & $\begin{pmatrix} 0 \\ 0 \end{pmatrix}$ \\
\hline
$P_0$ & $\begin{pmatrix} 0 & 0 \\ 0 & 0 \end{pmatrix}$ \\
\hline
$D$ & 0 \\
\hline
$R$ & 1 \\
\hline
\end{tabular}
\end{table}

\begin{figure}[htbp]
\centerline{\includegraphics[width=\columnwidth]{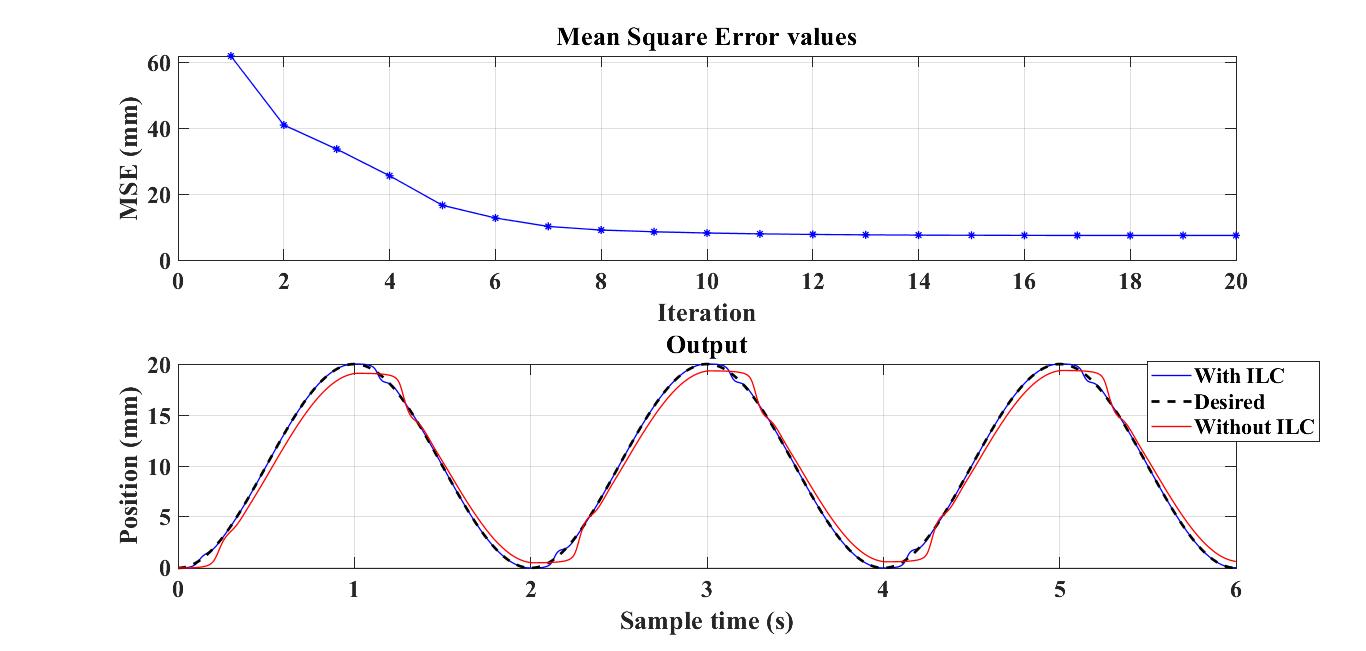}}
\caption{ILC-based friction compensation for a single task.\label{Fig.5}}
\end{figure}

\noindent
Following the simulation schematic in \textbf{Fig.~\ref{Fig.2}}, the plant response after 20 iterations using a sinusoidal reference of 0.5 Hz is shown in \textbf{Fig.~\ref{Fig.5}}. The dashed black line represents the desired trajectory, while the blue and red lines correspond to the plant output with and without ILC, respectively.

\noindent
ILC effectively compensates for friction effects in this scenario.  

\begin{figure}[htbp]
\centerline{\includegraphics[width=\columnwidth]{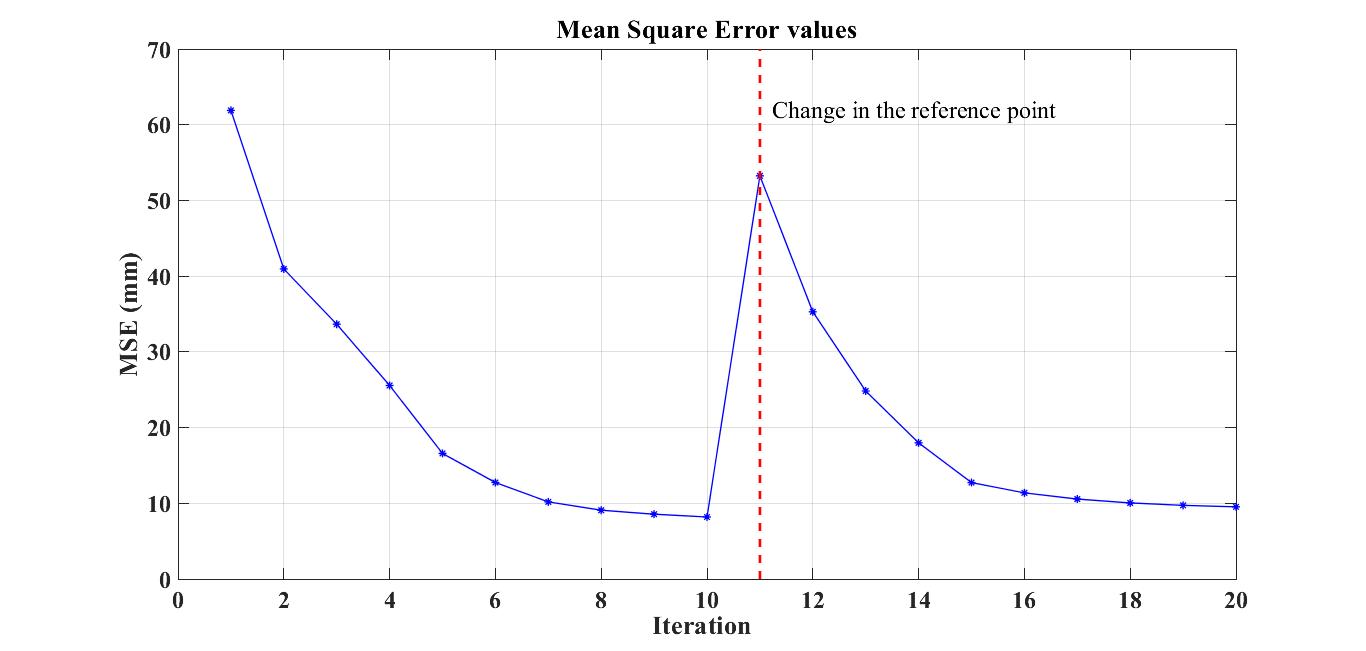}}
\caption{MSE evolution when the reference command is altered.\label{Fig.6}}
\end{figure}

\noindent
In a second scenario, the reference frequency is changed from 0.5 Hz to 0.6 Hz after 10 iterations. As shown in \textbf{Fig.~\ref{Fig.6}}, this causes an increase in mean square error (MSE) due to the change in friction patterns. The previously converged ILC effort becomes suboptimal, requiring additional iterations to adapt to the new reference.  

\noindent
To address this, the converged ILC efforts for multiple reference commands are used to train a simple lateral neural network. This network helps reduce the number of iterations needed to achieve the desired performance for new commands, while the ILC continues to compensate for residual disturbances and unmodeled dynamics, ensuring robust control across varying reference trajectories.

\section{Experimental Tests and Results}\label{sec5}

In the experimental setup, the actuator's rotor position is measured using a linear potentiometer. The analog signal is digitized via the STM32 microcontroller’s 16-bit ADC, providing high-resolution voltage detection over a 0–3.3 V range with a 400 mm stroke. Data is transmitted to a computer through UART at 1,000,000 bits per second. A software Interrupt Service Routine (ISR) triggers every 0.001 seconds to process the measured signal and generate the voltage command sent via a DAC pin to the servo amplifier.

\begin{figure}[htbp]
\centerline{\includegraphics[width=\columnwidth]{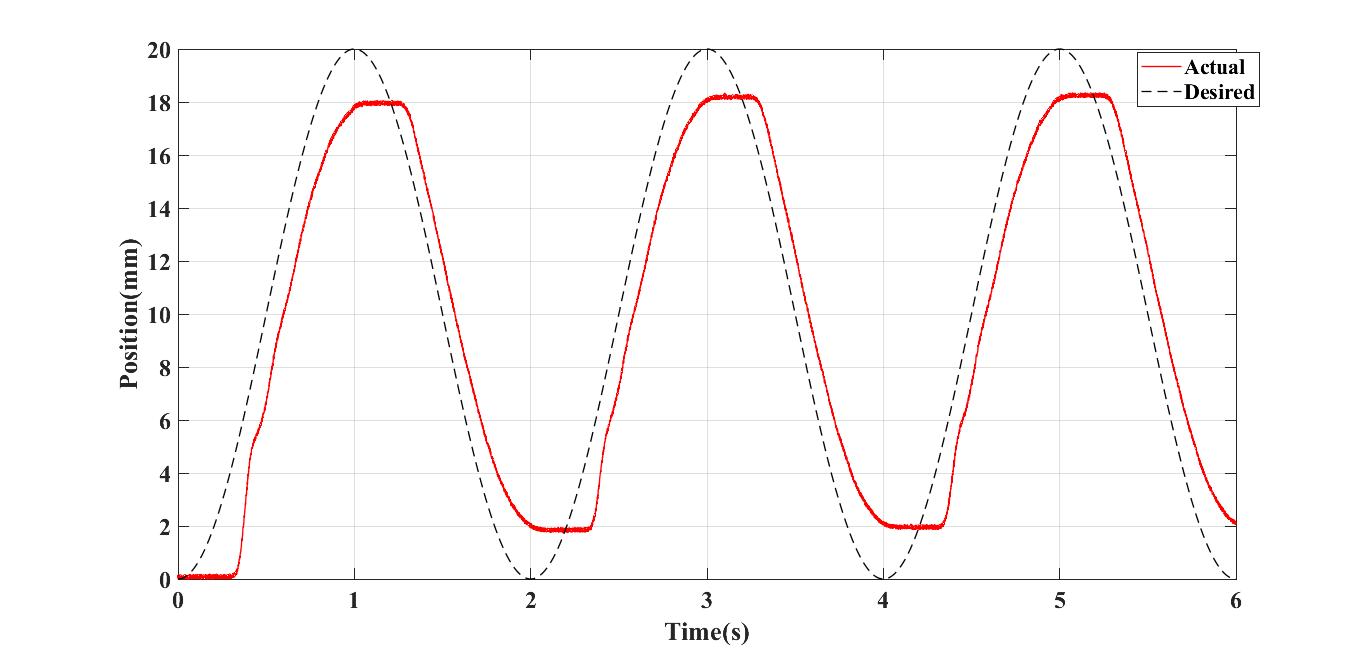}}
\caption{Trajectory tracking under a feedback controller, showing friction effects.\label{Fig.7}}
\end{figure}

\noindent
After implementing a basic feedback controller, the plant's response to a sinusoidal reference signal revealed the impact of friction, particularly at low frequencies, as illustrated in \textbf{Fig.~\ref{Fig.7}}.

\begin{figure}[htbp]
\centerline{\includegraphics[width=\columnwidth]{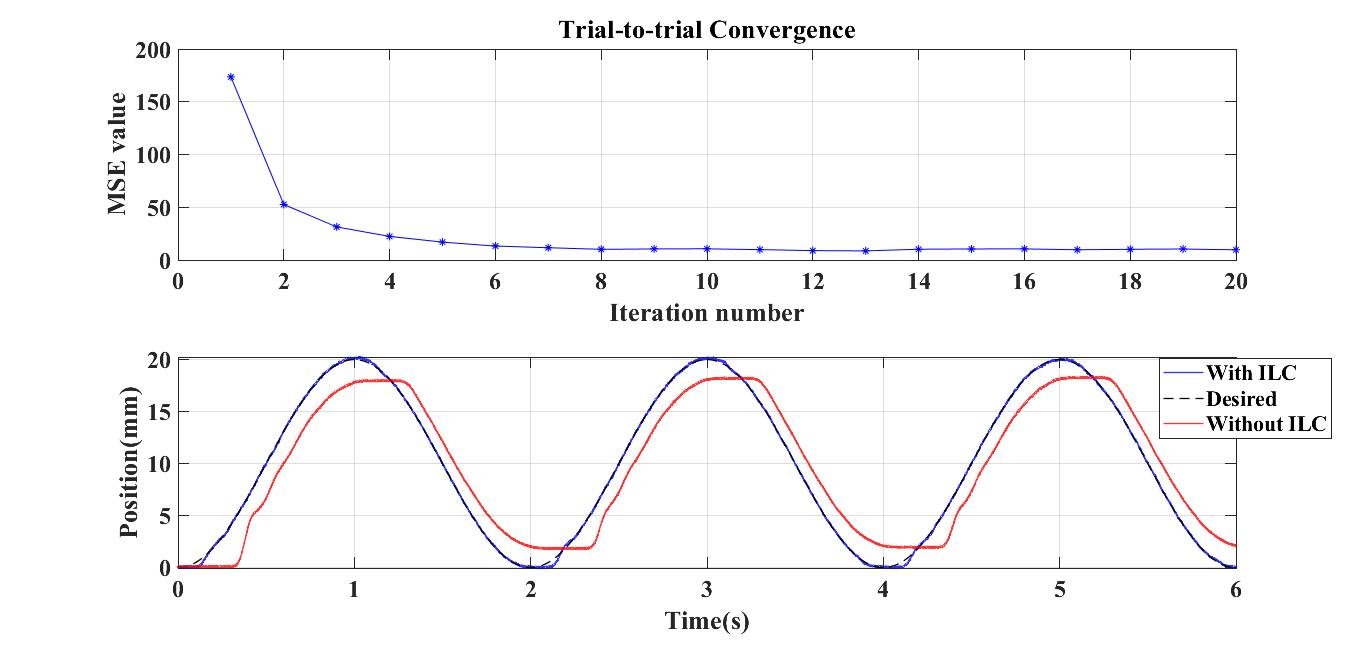}}
\caption{Friction compensation for a single task using ILC.\label{Fig.8}}
\end{figure}

\noindent
\textbf{Fig.~\ref{Fig.8}} demonstrates the performance improvement achieved with plant inversion ILC for a single reference task.

\begin{figure}[htbp]
\centerline{\includegraphics[width=\columnwidth]{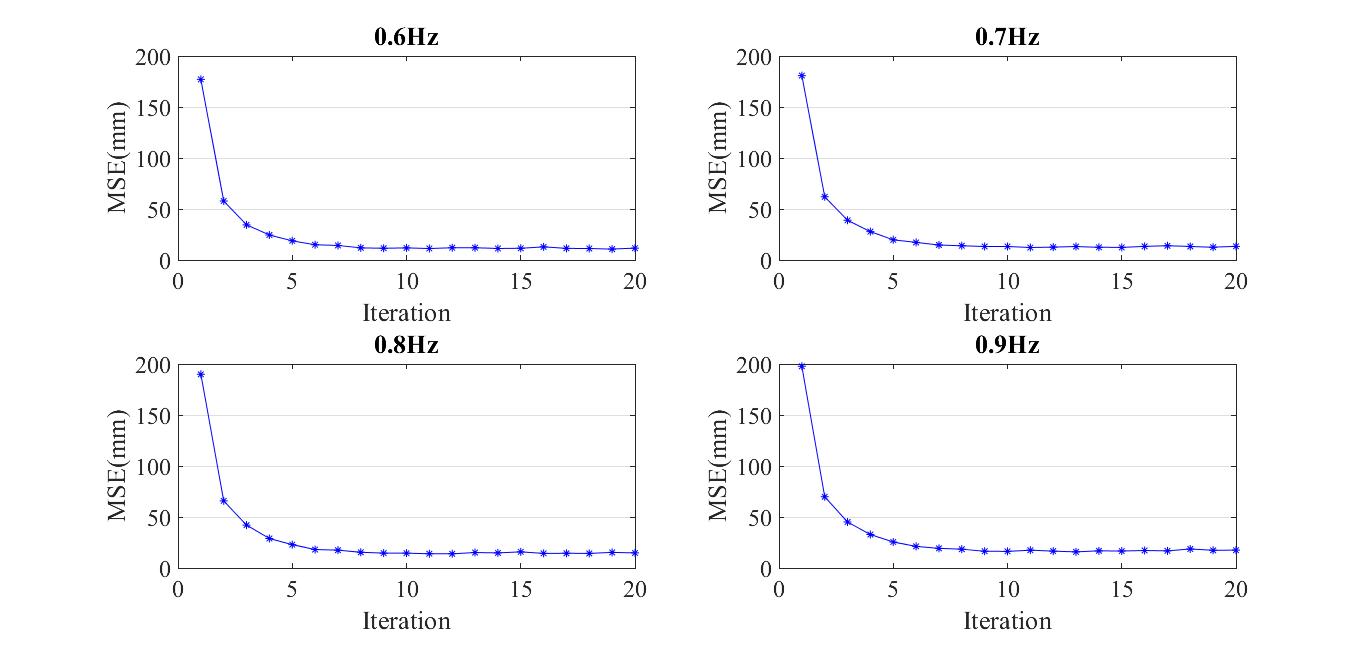}}
\caption{ILC performance across multiple reference commands.\label{Fig.9}}
\end{figure}

\noindent
Despite simplified model assumptions, such as constant force and structural rigidity, ILC effectively mitigates uncertainties. Using a Kalman filter allowed increasing the Q-filter bandwidth to 70 Hz, enhancing steady-state performance. The process was repeated for 29 additional reference commands, each introducing a new persistent uncertainty. \textbf{Fig.~\ref{Fig.9}} shows the MSE evolution over iterations for reference frequencies of 0.6 Hz, 0.7 Hz, 0.8 Hz, and 0.9 Hz.

\noindent
To address nonlinear effects from friction and model uncertainties, a neural network (illustrated in \textbf{Fig.~\ref{Fig.3}}) was trained using TensorFlow. The network predicts the nonlinear ILC effort based on the desired reference and velocity profile, capturing position-dependent friction variations~\cite{Norrlöf2020}.  

\noindent
The network architecture consists of three hidden layers with 8, 16, and 8 neurons. ReLU activations are used in the first two layers to capture nonlinearity, while the output layer uses a linear activation. Input data were normalized to ensure balanced contribution of all features. The model was compiled with the Adam optimizer and trained using the mean squared error loss over 100 epochs with a batch size of 128, improving stability and noise robustness.

\begin{figure}[htbp]
\centerline{\includegraphics[width=\columnwidth]{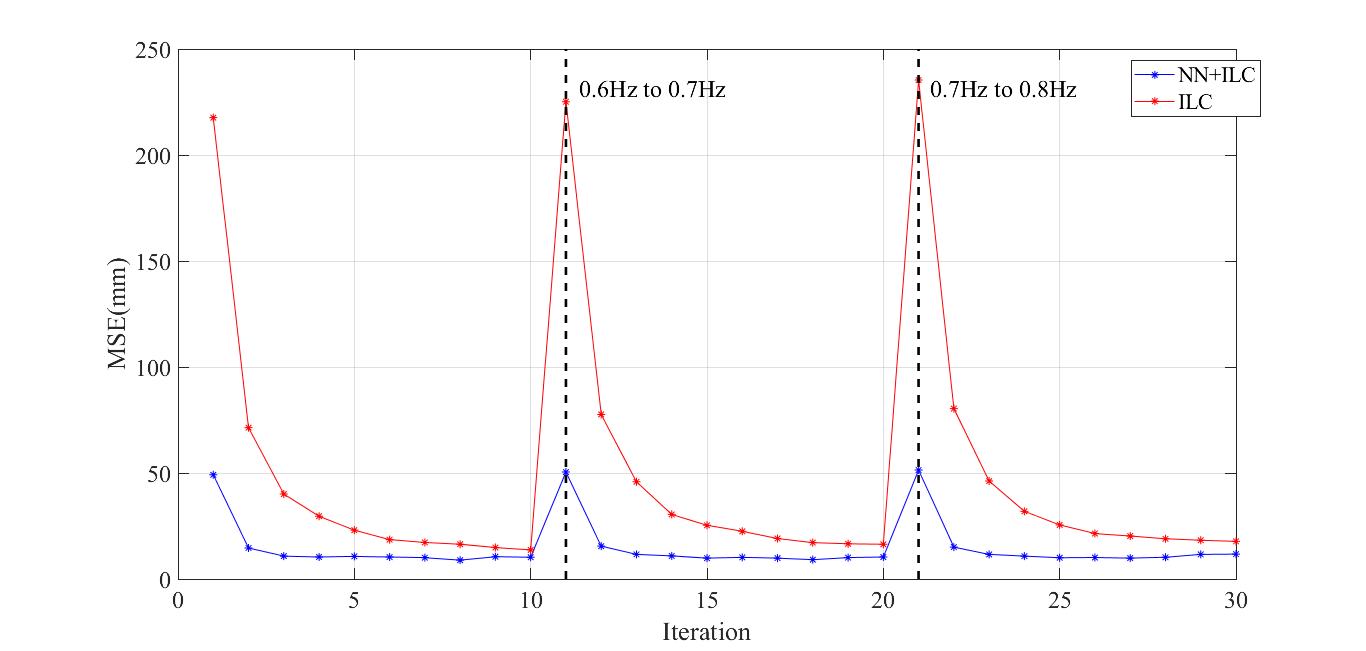}}
\caption{Comparison of convergence between conventional ILC and the proposed neural network-augmented ILC for multi-task scenarios.\label{Fig.10}}
\end{figure}

\noindent
After configuring the system as in \textbf{Fig.~\ref{Fig.3}}, the performance was evaluated over 30 iterations. The reference command was sequentially changed from 0.6 Hz to 0.7 Hz, and then 0.8 Hz every 10 iterations. \textbf{Fig.~\ref{Fig.10}} compares the convergence of the proposed neural network-augmented ILC with conventional ILC in a multi-task scenario.

\noindent
Results show that the initial MSE upon reference changes is significantly lower with the proposed method, and convergence is achieved more rapidly. The reduced uncertainty levels contribute to improved robustness, demonstrating the effectiveness of neural network-augmented ILC in compensating nonlinear effects and enhancing performance in multi-task applications.

\section{Conclusion}\label{sec6}

This study presents a novel approach that integrates Iterative Learning Control (ILC) with a simple lateral neural network to mitigate friction and uncertainties in Lorentz force actuators. The method combines the iterative performance improvement of ILC with the neural network's ability to estimate nonlinear friction effects and model mismatches. This hybrid strategy enables real-time adaptation, reducing the number of iterations required for convergence and minimizing extensive tuning.

\noindent
Experimental results demonstrate the effectiveness of the proposed approach, showing significant reductions in mean square error and faster convergence compared to conventional ILC. By employing a structurally simple neural network, the method remains practical for real-time applications, while maintaining robust performance across varying reference trajectories and friction conditions. Overall, this neural network-augmented ILC offers a promising solution for precise and efficient motion control in repetitive tasks.

\bibliography{sample}

\end{document}